\documentclass[12pt]{article}
\usepackage{epsfig}
\newlength{\defbaselineskip}
\newcommand{\setlinespacing}[1]%
           {\setlength{\baselineskip}{#1 \defbaselineskip}}

\begin{document}

\title{The systematic study of the influence of neutron excess on the fusion cross sections using different proximity-type potentials}

\author{O. N. Ghodsi and R. Gharaei\thanks{email: r.gharaei@stu.umz.ac.ir} \\
\\
{\small {\em  Sciences Faculty, Department of Physics, University of Mazandaran}}\\
{\small {\em P. O. Box 47415-416, Babolsar, Iran}}\\
}
\date{}
\maketitle

\begin{abstract}
\noindent Using different types of proximity potentials, we have
examined the trend of variations of barrier characteristics (barrier
height and its position) as well as fusion cross sections for 50
isotopic systems including various collisions of C, O, Mg, Si, S,
Ca, Ar, Ti and Ni nuclei with $1\leq N/Z < 1.6$ condition for
compound systems. The results of our studies reveal that the
relationships between increase of barrier positions and decrease of
barrier heights are both linear with increase of $N/Z$ ratio.
Moreover, fusion cross sections also enhance linearly with increase
of this ratio.

\end{abstract}

%======================================================================================
\newpage
\setlinespacing{1}
{\noindent \bf{1. INTRODUCTION}}\\

In recent years, using the neutron-rich projectiles and the
formation of the neutron-rich heavy nuclei, the interesting
properties of these fusion reactions have been discovered. In
general, the compound nucleus resulting from the fusion of the
neutron-rich nuclei is placed far from the $\beta$-stability line.
Furthermore, increasing the neutrons in the interaction nuclei leads
to reduction fusion barrier height. Therefore, fusion cross section
increases for neutron-rich projectile nuclei, with respect to nuclei
that lying near the stability line [1, 2].

For the first time, R. K. Puri \textit{et al.} performed a
systematic study on the isotopic dependence of fusion cross section
[3]. They calculated nuclear potential using the Skyrme energy
density model (SEDM), which it is quite successful in explaining the
fusion of two interaction nuclei at low energies [4-14], for
different colliding systems involving $^{A_{1}}$Ca+$^{A_{2}}$Ca,
$^{A_{1}}$Ca+$^{A_{2}}$Ni and $^{A_{1}}$Ni+$^{A_{2}}$Ni with $1\leq
N/Z \leq2$. In this condition, N and Z is neutron and proton numbers
of compound nucleus in each reaction. According to Ref. [3], with
increasing neutron in each group of these colliding systems, the
barrier heights $V_B$ decrease and the fusion cross sections
$\sigma_{fus}$ increase. Moreover, the variations of $R_B$, $V_B$
and $\sigma_{fus}$ with increasing ratio (N/Z) are linear (see Eqs.
(11), (12) and (15) of Ref. [3]).

Recently, R. K. Puri and N. K. Dhiman have also carried out two
systematic studies on the isotopic dependence of fusion
probabilities using SEDM and several different theoretical models
such as parameterized potentials due to Christensen and Winther,
$\textmd{Ng}\hat{\textmd{o}}$ and $\textmd{Ng}\hat{\textmd{o}}$ and
etc. Their investigations is consist of two ranges of different
isotopes collision of Ca and Ni nuclei with $0.6\leq N/Z \leq1$ [15]
and Ca, Ni, Ti and Ni nuclei with $0.5\leq N/Z \leq1$ [16]. Their
results show that the variations of heights and positions of barrier
are non-linear (second order), whereas the fusion cross sections
vary as linear with increasing ratio of $N/Z$ for colliding
different systems (for example, see Eqs. (11) and (12) of Ref.
[15]). The study of neutron rich nuclei is also reported at
heavy-ion collisions with intermediate energies. The effects of
isospin degree of freedom in collective and elliptic flow have been
studied, for example, in Refs. [17-20].

In this paper, we have performed a systematic study on the
relationships between variations of heights and positions of barrier
and fusion cross sections with increasing $N/Z$ ratio by using the
different proximity-type potentials, which shall be introduced in
the following section. We have selected thirteen groups of isotopic
systems, namely $^{A_{1}}$C+$^{A_{2}}$Si, $^{A_{1}}$O+$^{A_{2}}$Mg,
$^{A_{1}}$O+$^{A_{2}}$Si, $^{A_{1}}$Si+$^{A_{2}}$Si,
$^{A_{1}}$Mg+$^{A_{2}}$S, $^{A_{1}}$Si+$^{A_{2}}$Ni,
$^{A_{1}}$Ca+$^{A_{2}}$Ca, $^{A_{1}}$S+$^{A_{2}}$Ni,
$^{A_{1}}$Ar+$^{A_{2}}$Ni, $^{A_{1}}$Ca+$^{A_{2}}$Ni,
$^{A_{1}}$Ni+$^{A_{2}}$Ni, $^{A_{1}}$Ti+$^{A_{2}}$Ni and
$^{A_{1}}$Ca+$^{A_{2}}$Ti with $1\leq N/Z < 1.6$. In all these
systems, the empirical data have been reported (see Table 1). In
this work, we follow the procedure proposed in Ref. [3].
\\
\\
\\
%======================================================================================

\noindent{\bf {2. THE PROXIMITY FORMALISM}}\\

The interaction potential between target and projectile nuclei is
one of the most important factors in the description of the fusion
reaction. In general, this potential consists of two parts,
short-range nuclear attraction $V_{N}(r)$ and large-range coulomb
repulsion $V_{C}(r)$. Proximity model is one of the practical types
in calculation of nuclear potential. When two surfaces are
approaching each other, approximately in distance of 2-3 fm, an
additional force due to the proximity of the surface will appear
which is called as proximity potential. All versions of this model
are based on the proximity force theorem [21]. According to this
theory, nuclear part of total interaction potential is product of a
factor depending on the mean curvature of the interaction surface
and a universal function depending on the separation distance.
Furthermore, nuclear potentials in the proximity formalism are
independent of the masses of interaction nuclei. Using the proximity
potentials dependence on the liquid drop model, one expects that the
nuclear matter incompressibility is one of the intrinsic properties
in this formalism.

12 different versions of the proximity model have been introduced by
R. K. Puri \textit{et al}. in Ref. [22]. Their investigations show
that all proximity potentials are able to reproduce experimental
data within ±10\%, on the average. Among various potentials in Ref.
[22], the Aage Winther (AW 95) [23], Bass 1980 (Bass 80) [24] and
Denisov Potential [25] (Denisov DP, which in this paper we have used
as Prox. DP) have the best agreement with the experimental data.
Therefore, in first step, we have selected these versions for
calculating the nuclear potential. In addition to above potentials,
we have used the Proximity 2010 (Prox. 2010) potential [26] for
study of the isotopic dependence in all thirteen mentioned systems.
In the following subsections, we have briefly introduced these
versions of proximity potential.
\\
\\
\\
\noindent{\bf {A. Aage Winther (AW 95)}}\\

Aage Winther, in 1995, introduced a form of nuclear potential by
taking Woods-Saxon parametrization as [23],

\begin{equation} \label{1}
V_{N}(r)=-\frac{V_{0}}{1+ \textmd{exp} (\frac{r-R_1-R_2}{a})},
\end{equation}
with

\begin{equation} \label{2}
V_{0}=16\pi\frac{R_{1}R_{2}}{R_1+R_2}\gamma a,
\end{equation}
here surface energy coefficient $\gamma$ has the following form

\begin{equation} \label{3}
\gamma=0.95\bigg[1-1.8\bigg(\frac{N_P-Z_P}{A_P}\bigg)\bigg(\frac{N_T-Z_T}{A_T}\bigg)\bigg].
\end{equation}
Winther adjusted the diffuseness $\textit{a}$ and radius $R_{i}$
parameters through a wide comparison with experimental data for
heavy-ion elastic scattering. This refined adjustment leads to a new
form of $\textit{a}$ and $R_{i}$ parameters, which have been defined
by Eqs. (8) and (9) of Ref. [23].
\\
\\
\\
\noindent{\bf {B. Bass 1980 (Bass 80)}}\\

Bass in 1977 [27] and 1980 [28] introduced a nucleus-nucleus
potential which is labeled as Bass 80. This potential has been
derived from the liquid drop model and geometric interpretation of
the fusion data in the above regions of the barrier for systems with
$Z_{P}Z_{T}=64-850$. In this model the nuclear part of the
interaction potential can be written as [24],

\begin{equation} \label{4}
V_{N}(r)=\frac{R_{1}R_{2}}{R_1+R_2}\Phi(s),
\end{equation}
where $s=r-R_1-R_2$. Here the universal function $\Phi(s)$ has the
following form,

\begin{equation} \label{5}
\Phi(s)=\bigg[0.033\textmd{exp}(\frac{s}{3.5})+0.007\textmd{exp}
(\frac{s}{0.65})\bigg]^{-1}
\end{equation}
with central radius, $R_{i}$, as

\begin{equation} \label{6}
R_{i}=R_{s}\bigg(1-\frac{0.98}{R^{2}_{s}}\bigg)\qquad (i=1, 2)
\end{equation}
where the sharp radii $R_{s}$ has been defined by using the Eq. (4)
of Ref. [22].
\\
\\
\\
\noindent{\bf {C. Denisov DP (Prox. DP)}}\\

Denisov, using the semi-microscopic approximation and examination of
119 spherical or near spherical even-even nuclei around the
$\beta$-stability line, has defined an expression for nuclear part
of total interaction potential as the following form [25],

\begin{equation} \label{7}
V_{N}(r)=-1.989843\overline{R}\Phi(s)\newline
\times\bigg[1+0.003525139(\frac{A_1}{A_2}+\frac{A_2}{A_1})^{{3}/{2}}-0.4113263(I_1+I_2)\bigg].
\end{equation}
where $s=r-R_1-R_2-2.65$ and $\overline{R}=R_{1}R_{2}/(R_1+R_2)$.
The explicit form of the universal function $\Phi(s)$ and effective
nuclear radius $R_{i}$ have been defined by Eqs. (11), (12) and (13)
of Ref. [25].
\\
\\
\\
\noindent{\bf {D. Proximity 2010 (Prox. 2010)}}\\

The effect of surface energy coefficient $\gamma$ in the proximity
potential, has been discussed in Ref. [26]. In this study, four
different versions of surface energy coefficient $\gamma$ have been
introduced based on the Prox. 77 potential [21]. The obtained
results show that the barrier heights and positions as well as
fusion cross sections due to Prox. 77 potential with new surface
energy coefficient, namely $\gamma$-MN1976, have the best agreement
with experimental data (see Figs. (3) and (4) of Ref. [26]). This
modified proximity potential is labeled as proximity 2010. In this
model, the nuclear potential $V_{N}(r)$ has the following form,

\begin{equation} \label{8}
V_{N}(r)=4\pi\overline{R}\gamma\Phi(r-C_1-C_2),
\end{equation}
with

\begin{equation} \label{9}
\gamma=\gamma_{0}\bigg[1-K_{s}\bigg(\frac{N-Z}{A}\bigg)^2\bigg].
\end{equation}

Among different values for constants $\gamma_0$ and $K_{s}$, the set
of $\gamma_0=1.460734$ $\textmd{MeV}/\textmd{fm}^2$ and $K_{s}= 4.0$
have the best results for barrier heights and fusion cross sections
in the many systems that have been evaluated in Ref. [26]. In Eq.
(8), $\overline{R}$ is the reduced radius and $\textmd{C}_1$ (or
$\textmd{C}_2$) is known as S$\ddot{\textmd{u}}$ssmann$^{'}$s
centarl radius. The explicit form of these parameters as given by
Eqs. (2) and (3) of Ref. [22]. The universal function
$\Phi(r-C_1-C_2)$ has been defined by Eq. (6) of Ref. [22].
\\
\\
\\
%==========================================================================================================
\noindent{\bf {3. CALCULATIONS}}\\

In order to calculate the nuclear potential, we have employed four
versions of the proximity model, \textit{i.e.} AW 95, Bass 80, Prox.
DP and Prox. 2010 potentials. By adding the Coulomb potential to a
nuclear part, we get the total potential $V_{T}(r)$ as,

\begin{equation} \label{10}
V_{tot}(r)=V_{C}(r)+V_{N}(r),
\end{equation}
where $V_{C}(r)={Z_1 Z_2 e^2}/{r}$. On the other hand, one can
determine the barrier position $R_B$ and height $V_B$ by the
following conditions,

\begin{equation} \label{11}
{\bigg(\frac{dV_{tot}(r)}{dr}\bigg)}_{r=R_B}=0\qquad;\qquad{\bigg(\frac{d^2V_{tot}(r)}{dr^2}\bigg)}_{r=R_B}\leq{0}
\end{equation}
The calculated values of the $R_B$ and $V_B$ resulting from the AW
95, Bass 80, Prox. DP and Prox. 2010 potentials have been listed in
Table 1-4. In these tables, the values of the Coulomb $V_{C}(r)$ and
nuclear $V_{N}(r)$ potentials in $r=R_B$ have also been listed for
each reaction. As a result from this calculations, positions and
heights of barrier, respectively, increase and decrease with
addition of neutron in different colliding systems. The experimental
data for barrier heights and fusion cross sections are available for
systems with nuclei that are little far from the stability line.
However, using the proposed semi-empirical approach in Ref. [29],
one can calculate the cross sections for fusion reactions, but doing
this work requires the independent calculations which we will
employe it as a useful approach in the further investigations.
Although, all considered potentials are already applied to more than
400 reactions and compared with data [22, 26], in Fig. 1, the ratio
of experimental and calculated values of barrier height and position
as a function of $(N/Z-1)$ for different versions of proximity
potentials are plotted. These values, for example, have been
calculated for $^{40}$Ca+$^{40, 48}$Ca, $^{40}$Ca+$^{58, 62}$Ni,
$^{28}$Si+$^{28, 30}$Si, $^{30}$Si+$^{28}$Si, $^{58}$Ni+$^{58,
64}$Ni, $^{64}$Ni+$^{64}$Ni, $^{24}$Mg+$^{32, 34}$S and
$^{26}$Mg+$^{32, 34}$S systems. As can be seen from Fig. 1, our
results for $R_B$ and $V_B$ are in good agreement with experimental
data for all proximity versions.

For systematic study of the variations of $R_B$ and $V_B$, the
percentage difference of the heights and positions of the fusion
barrier with respect to their corresponding values for the N=Z cases
defined as,

\begin{equation} \label{12}
\Delta{R_B(\%)}=\frac{R_B-R_B^0}{R_B^0}\times100,
\end{equation}

\begin{equation} \label{13}
\Delta{V_B(\%)}=\frac{V_B-V_B^0}{V_B^0}\times100,
\end{equation}
where $R_B^0$ and $V_B^0$ are the positions and heights of the
barrier for the $N=Z$ cases (As earlier stated, $N=N_1+N_2$ and
$Z=Z_1+Z_2$; $N_i$ and $Z_i$ are the neutron and proton numbers of
two interaction nuclei, respectively). In all systems, where the
values of $R_B^0$ and $V_B^0$ are not available, a straight-line
interpolation is used between the known points to compute the
$\Delta{R_B(\%)}$ and $\Delta{V_B(\%)}$. The obtained results for
percentage difference of ${R_B}$ and ${V_B}$ for all selected
versions of proximity potential, are plotted in Fig. 2. The results
of other investigations have also been displayed in this figure,
which show good agreement with our obtained results. As one can see
in Fig. 2, the increase of barrier positions and decrease of barrier
heights are both linear. We have parameterized these processes by
following relations,

\begin{equation} \label{14}
\Delta{R_B(\%)}=\alpha\bigg(\frac{N}{Z}-1\bigg),
\end{equation}

\begin{equation} \label{15}
\Delta{V_B(\%)}=\beta\bigg(\frac{N}{Z}-1\bigg),
\end{equation}
whose the values of the constants $\alpha$ and $\beta$ for AW 95,
Bass 80, Prox. DP, and Prox. 2010 potentials have been listed in
Table 5. We expect that with increase of neutron in the interaction
nuclei of thirteen groups of colliding systems, the nuclear
attractive force increases and therefore the heights of barrier
decrease (see Fig. 2). To display the increase of nuclear potential
with addition of neutron, like Eqs. (14) and (15), we have
parameterized the trend of $V_N$ variations. The obtained results
have been shown in Fig. 3. As one can see from this figure, the
nuclear potential, like height and position of barrier, increase
linearly. This process has been parameterized by following
relations,

\begin{equation} \label{16}
\Delta{V_N(\%)}=\gamma\bigg(\frac{N}{Z}-1\bigg),
\end{equation}
whose the values of the constants $\gamma$ for our selected
potentials have also been listed in Table 5.

One can also examine this phenomenon by a different approach.
According to the Pauli exclusion principle, which prevent the
overlapping of the wave functions of two systems of fermions, we
expect the interaction potential between two colliding nuclei will
contain an additional repulsive interaction. In nuclear fusion
process, when two interaction nuclei complete overlap the nuclear
matter density is twice that of the saturation case,
$\rho\approx2\rho_0$. This conduct of density distributions have
been displayed, for example, for $^{40}$Ca + $^{40}$Ca fusion
reaction in Fig. 4. In this figure, the solid curves are based on
the density distributions of the target and projectile nuclei,
whereas the short-dashed curves are based on the total density
distribution. It is clear from Fig. 4(c) that the total density at
complete overlap of interaction nuclei almost becomes twice its
initial value.

According to the nuclear equation of state, where energy per nucleon
is proportional to density, increasing density in the overlapping
region of two interacting nuclei leads to an increase in the energy
of the compound system. This increase of the energy can be
attributed to the short-range repulsive interaction in the nuclear
part of total interaction potential. By modeling the repulsive core
effects [30], shallow packet appears in the inner part of the
barrier (see, for example, Fig. 2 of Ref. [30]). The effect of
nuclear matter incompressibility on the inner regions of the fusion
barrier and the depth of pocket is essential. As a result, a soft
nuclear matter provides a deep pocket , whereas a hard one provides
a shallow pocket. With the addition of neutrons in each of the
thirteen interaction systems, we expect that the nuclear matter
increases in the overlap region of the density distributions. This
phenomenon leads to an increase of energy and consequently increase
of repulsive force resulting from nuclear matter incompressibility
effects. This additional repulsive force increases the barrier
height and reduces the potential depth.

In Fig. 5, for example, we have shown the total interaction
potential by using the two versions of proximity potential, AW 95
and Prox. DP, for $^{A_{1}}$Ni+$^{A_{2}}$Ni system. As can be seen
in Fig. 5, with increasing neutron in fusion reactions, the pocket
energy $V_{pocket}$ is reduced. Therefore, it is predictable that
the increase of attractive force could be dominate the increase of
the repulsive force.

One-dimensional barrier penetration is one of the applied models for
calculation of the fusion cross section. In this formalism, the
fusion cross section is given by,

\begin{equation} \label{17}
\sigma_{fus}=\frac{\pi\hbar^2}{2\mu{E_{c.m.}}}\sum^{l_{max}}_{l=0}(2l+1)T_l(E_{c.m.})
\end{equation}
where $T_l(E_{c.m.})$ is the quantum-mechanical transmission
probability through the potential barrier for the $l$-$th$ partial
wave and $\mu$ is reduced mass of the target and projectile system.
With assumption that the width and position of the barrier are
independent of angular momentum $l$ and with $E_{c.m.}\gg{V_B}$, the
Eq. (17) is reduced to sharp cutoff formula,

\begin{equation} \label{18}
\sigma_{fus}(mb)=10\pi{R^2_B}\bigg(1-\frac{V_B}{E_{c.m.}}\bigg).
\end{equation}

In Fig. 6, we have displayed the fusion cross sections, Eq. (18),
for $^{40}$Ca+$^{58,60,62}$Ni systems using all four proximity
potentials. In this figure, the dashed, short-dashed and dash-dotted
curves based on the calculated fusion cross sections for
$^{40}$Ca+$^{58}$Ni, $^{40}$Ca+$^{60}$Ni  and $^{40}$Ca+$^{62}$Ni,
respectively. One observes that the obtained results have a good
agreement with experimental data [31], in above barrier. Moreover,
with addition of neutron in these fusion reactions, the fusion
probabilities are increased.

The percentage difference for fusion cross section is given by the
following relation,

\begin{equation} \label{19}
\Delta{\sigma_{fus}(\%)}=\frac{\sigma_{fus}(E_{c.m.}^0)-\sigma^{0}_{fus}(E_{c.m.}^0)}{\sigma^{0}_{fus}(E_{c.m.}^0)}\times100,
\end{equation}
where $E_{c.m.}^0={E_{c.m.}}/{V_B^0}$. According to the condition
$E_{c.m.}\gg{V_B}$, we have calculated this percentage difference
for two center-of-mass energies $E_{c.m.}=1.125V_B^0$ and
$E_{c.m.}=1.375V_B^0$, for example. The obtained results for AW 95,
Bass 80, Prox. DP, and Prox. 2010 potentials have been shown in Fig.
7. As see from this figure, the relationship between changes of the
fusion cross sections with increasing neutron (or ratio $N/Z$) in
four versions of potential are linear. This relation is given by,

\begin{equation} \label{20}
\Delta{\sigma_{fus}(\%)}=c\bigg(\frac{N}{Z}-1\bigg),
\end{equation}
where the values of constant c for different potentials and energies
have been listed in Table 5. In order to reduce the barrier height
with increasing neutron, see Table 1-4, one expects an increase of
fusion cross section in each of the interaction systems. This
phenomenon is well visible in Fig. 7.
\\
\\
\\
%==========================================================================================================
\noindent{\bf {4. CONCLUSION}}\\

Our purpose in this paper is a systematic study on the neutron
excess effect in 50 isotopic reactions $^{A_{1}}$C+$^{A_{2}}$Si,
$^{A_{1}}$O+$^{A_{2}}$Mg, $^{A_{1}}$O+$^{A_{2}}$Si,
$^{A_{1}}$Si+$^{A_{2}}$Si, $^{A_{1}}$Mg+$^{A_{2}}$S,
$^{A_{1}}$Si+$^{A_{2}}$Ni, $^{A_{1}}$Ca+$^{A_{2}}$Ca,
$^{A_{1}}$S+$^{A_{2}}$Ni, $^{A_{1}}$Ar+$^{A_{2}}$Ni,
$^{A_{1}}$Ca+$^{A_{2}}$Ni, $^{A_{1}}$Ni+$^{A_{2}}$Ni,
$^{A_{1}}$Ti+$^{A_{2}}$Ni and $^{A_{1}}$Ca+$^{A_{2}}$Ti with $1\leq
N/Z < 1.6$. The nuclear part of total interaction potential has been
calculated by using the proximity potentials AW 95, Bass 80, Prox.
DP, and Prox. 2010, whose values of heights and positions of the
barrier and fusion cross sections in these potentials according to
Refs. [22, 26, 32], have the best agreement with experimental data.
In this work, the obtained results are included the following cases:

(i) We have found a linear relation between changes of the barrier
position $\Delta{R_B}$ and barrier height $\Delta{V_B}$ with
increasing of the ratio $N/Z$ in thirteen groups of the fusion
reactions (see Eqs. (14) and (15)). (ii) The parametrization of
nuclear potential $V_{N}(r)$ in $r=R_B$ shows that the trend of
$V_{N}(r)$ versus $N/Z$ ratio is linear for all considered
potentials. (iii) The changes of the fusion cross sections
$\Delta\sigma_{fus}$ with increasing of the ratio $N/Z$ in thirteen
groups of the interaction systems are linear (see Eq. (20)). (iv)
The reduction the Coulomb barrier height and consequently increase
of the fusion cross section has been attributed to increase of the
attractive force due to neutron excess. (v) The attractive force
resulting from adding neutrons can be overcome on the repulsive
force due to the incompressibility effects.

As a further investigation, one can examine the addition of neutron
effects using a systematic study based on the dynamic approach. Ning
Wang \textit{et al.}, using improved quantum molecular dynamics
(QMD) model, have discussed the influence of dynamic corrections in
the near coulomb barrier [33]. The values of Coulomb barrier height
and the depth of the mean potential well of the compound nuclei have
been calculated in fusion $^{40}$Ca+$^{90}$,$^{96}$Zr and
$^{48}$Ca+$^{90}$Zr at energies $E_{c.m.}= 95.0$ MeV (below the
barrier) and $E_{c.m.}= 107.6$ MeV (above the barrier) (see Table
III of Ref. [33]). The obtained results show that dynamic effects
decrease height and thickness of barrier. All these problems are
very important for further understanding the mechanism of neutron or
proton rich nuclei. We are going to study these aspects in future
works.
\\
\\
\\
%=======================================================================================================================================

%========================================================================================================
\newpage
\noindent{\bf {FIGURE CAPTIONS}}\\
\\
Fig. 1. The ratio of experimental [36-38,41,42,44,52] and calculated
values of barrier height and position based on the AW 95, Bass 80,
Prox. DP, and Prox. 2010 potentials for the fusion reactions of
$^{24}$Mg+$^{32, 34}$S, $^{26}$Mg+$^{32, 34}$S, $^{28}$Si+$^{28,
30}$Si, $^{30}$Si+$^{28}$Si, $^{40}$Ca+$^{40, 48}$Ca,
$^{40}$Ca+$^{58, 62}$Ni, $^{58}$Ni+$^{58, 64}$Ni and
$^{64}$Ni+$^{64}$Ni, for example, as a function of $N/Z-1$.
\\
\\
Fig. 2. The percentage increase of fusion barrier position $R_B$
(left panels) and percentage decrease of fusion barrier height $V_B$
(right panels) with respect to their corresponding values for $N=Z$
cases, as a function of the ratio $N/Z$ of the compound system for
the AW 95, Bass 80, Prox. DP, and Prox. 2010 potentials. The solid
line in each graph is the result of the linear fitting to the
calculated values of $\Delta{R_B}(\%)$ and $\Delta{V_B}(\%)$. The
results of other model have also been displayed [16, 34].
\\
\\
Fig. 3. The percentage increase of nuclear potential $V_N$ with
respect to its corresponding values for $N=Z$ cases, as a function
of the ratio $N/Z$ of the compound system for the AW 95, Bass 80,
Prox. DP, and Prox. 2010 potentials. The solid line in each graph is
the result of the linear fitting to the calculated values of
$\Delta{V_N}(\%)$.
\\
\\
Fig. 4. The process of density distributions overlap of interaction
nuclei for $^{40}$Ca+$^{40}$Ca reaction in (a) $\rho\approx\rho_0$,
(b) $\rho_0<\rho<2\rho_0$ and (c) $\rho\approx2\rho_0$, which $\rho$
and $\rho_0$ are density in the overlapping region and saturation
density, respectively. Total density distribution is shown with
dotted curve.
\\
\\
Fig. 5. Ion-Ion potentials for $^{A_1}$Ni+$^{A_2}$Ni system based on
the (a) AW 95 and (b) Prox. DP potentials .  The pocket energy
$V_{pocket}$ is also indicated in one reaction.
\\
\\
Fig. 6. The fusion cross sections as a function of center of mass
energy for $^{40}$Ca+$^{58}$Ni, $^{40}$Ca+$^{60}$Ni and
$^{40}$Ca+$^{62}$Ni based on the (a) AW 95, (b) Bass 80, (c) Prox.
DP, and (d) Prox. 2010 potentials. The experimental data is taken
from the Ref. [31].
\\
\\
Fig. 7. The percentage increase of fusion cross section
$\sigma_{fus}$ for $E_{c.m.}$=1.125$V_B^0$ (left panels) and
$E_{c.m.}$=1.375$V_B^0$ (right panels) with respect to their
corresponding values for $N=Z$ cases, as a function of the ratio
$N/Z$ of the compound system for the (a) AW 95 (b) Bass 80 (c) Prox.
DP and (d) Prox. 2010 potentials. The solid line in each graph is
the result of the linear fitting to the calculated values of
$\Delta\sigma_{fus}(\%)$.
\\
\\
\newpage
\noindent{\bf {TABLE CAPTIONS}}
\\
Table 1. The calculated barrier positions $R_B$ and heights $V_B$
for different fusion reactions using AW 95 potential, compared with
the empirical data. The nuclear and Coulomb potentials in $r=R_B$
have also been reported. The systems are listed with respect to
their increasing $Z_{1}Z_{2}$ values.

\begin{center}
\begin{tabular}{c c c c c c c c c c}
  \hline
  \hline
  Reaction & N/Z & $Z_{1}Z_{2}$& $R_B$   &  $V_C$   &  $V_N$   &  $V_B$  & $R^{Emp.}_B$ & $V^{Emp.}_B$ & Ref.   \\
  \hline
  \cline{4-10}
  $^{12}$C+$^{28}$Si     &   1      & 84 &   8.47    &  14.28    &   -1.05    &   13.23   &  7.42$\pm$0.2  &  12.59$\pm$0.3  &  [37] \\
  $^{12}$C+$^{29}$Si     &   1.05   & 84 &   8.53    &  14.18    &   -1.03    &   13.15   &  8.19          &  13.46          &  [36] \\
  $^{12}$C+$^{30}$Si     &   1.1    & 84 &   8.58    &  14.10    &   -1.03    &   13.07   &  8.39          &  13.20          &  [36] \\
  \\
  $^{16}$O+$^{24}$Mg     &   1      & 96 &   8.52    &  16.22    &   -1.20    &   15.02   &  8.40$\pm$0.4  &  15.90$\pm$0.9  &  [38] \\
                         &          &    &           &           &            &           &  8.48          &  16.00          &  [39] \\
  $^{16}$O+$^{26}$Mg     &   1.1    & 96 &   8.65    &  15.98    &   -1.17    &   14.81   &  8.70$\pm$0.4  &  16.50$\pm$0.9  &  [38] \\
  $^{18}$O+$^{24}$Mg     &   1.1    & 96 &   8.70    &  15.89    &   -1.15    &   14.74   &  7.80$\pm$0.3  &  14.90$\pm$0.9  &  [40] \\
  \\
  $^{16}$O+$^{28}$Si     &   1      & 112 &  8.66    &  18.62    &   -1.36    &   17.26   &  7.98          &  17.23   &  [39] \\
  $^{16}$O+$^{29}$Si     &   1.045  & 112 &  8.72    &  18.49    &   -1.34    &   17.15   &  9.12          &  16.30   &  [36] \\
  $^{16}$O+$^{30}$Si     &   1.090  & 112 &  8.77    &  18.39    &   -1.33    &   17.05   &  9.18          &  16.12   &  [36] \\
  $^{18}$O+$^{28}$Si     &   1.090  & 112 &  8.83    &  18.26    &   -1.32    &   16.94   &  8.76          &  16.90   &  [36] \\
  \\
  $^{24}$Mg+$^{32}$S     &   1        & 192 &   9.08    &  30.45    &   -2.21    &   28.24   &  8.70$\pm$0.3  &  28.10$\pm$1.6  &  [38] \\
                         &            &     &           &           &            &           &  9.20          &  27.93          &  [41] \\
  $^{24}$Mg+$^{34}$S     &   1.071    & 192 &   9.18    &  30.12    &   -2.18    &   27.94   &  9.40          &  27.38          &  [41] \\
  $^{26}$Mg+$^{32}$S     &   1.071    & 192 &   9.21    &  30.02    &   -2.15    &   27.87   &  9.36          &  27.48          &  [41] \\
  $^{26}$Mg+$^{34}$S     &   1.143    & 192 &   9.31    &  29.70    &   -2.10    &   27.60   &  9.50          &  27.11          &  [41] \\
  \\
  $^{28}$Si+$^{28}$Si     &   1        & 196 &   9.09    &  31.05    &   -2.25    &   28.80   &  8.94          &  28.89         &  [42] \\
                          &            &     &           &           &            &           &  8.25$\pm$0.2  &  28.95$\pm$0.7 &  [37] \\
  $^{28}$Si+$^{30}$Si     &   1.071    & 196 &   9.20    &  30.68    &   -2.22    &   28.46   &  8.86          &  29.13         &  [42] \\
                          &            &     &           &           &            &           &  8.47$\pm$0.2  &  28.28$\pm$0.7 &  [37] \\
  $^{30}$Si+$^{30}$Si     &   1.143    & 196 &   9.31    &  30.32    &   -2.17    &   28.15   &  9.06          &  28.54         &  [42] \\
  \\
  $^{28}$Si+$^{58}$Ni     &   1.048   & 392 &   9.84     &  57.37    &   -4.07    &   53.30   &   9.00$\pm$0.9    &  53.80$\pm$0.8  &  [43] \\
  $^{28}$Si+$^{62}$Ni     &   1.143   & 392 &   9.98     &  56.57    &   -3.94    &   52.63   &   9.89            &  52.89          &  [43] \\
  $^{28}$Si+$^{64}$Ni     &   1.190   & 392 &   10.04    &  56.23    &   -3.92    &   52.31   &   9.20$\pm$1.0    &  52.40$\pm$1.1  &  [43] \\
  $^{30}$Si+$^{58}$Ni     &   1.095   & 392 &   9.96     &  56.68    &   -3.95    &   52.73   &   8.30$\pm$1.1    &  52.20$\pm$1.2  &  [43] \\
  $^{30}$Si+$^{62}$Ni     &   1.190   & 392 &   10.09     & 55.95    &   -3.85    &   52.10   &   9.70$\pm$1.0    &  52.20$\pm$0.9  &  [43] \\
  $^{30}$Si+$^{64}$Ni     &   1.238   & 392 &   10.15     & 55.61    &   -3.81    &   51.80   &   9.40$\pm$0.8    &  51.20$\pm$0.9  &  [43] \\
  \hline
\end{tabular}
\end{center}

\newpage
Table 1. (Continued)

\begin{center}
\begin{tabular}{c c c c c c c c c c}
  \hline
  \hline
  Reaction & N/Z & $Z_{1}Z_{2}$& $R_B$   &  $V_C$   &  $V_N$   &  $V_B$  & $R^{Emp.}_B$ & $V^{Emp.}_B$ & Ref.   \\
  \hline
  \cline{4-10}
  $^{40}$Ca+$^{40}$Ca     &   1     & 400 &   9.74     &  59.14    &   -4.22    &   54.92   &   9.50$\pm$0.5    &  50.60$\pm$2.8  &  [40] \\
                          &          &    &            &           &            &           &   8.80$\pm$0.5    &  52.30$\pm$0.5  &  [44] \\
                          &          &    &            &           &            &           &   9.10$\pm$0.6    &  55.60$\pm$0.8  &  [46] \\
  $^{40}$Ca+$^{44}$Ca     &   1.1   & 400 &   9.91     &  58.13    &   -4.12    &   54.01   &   8.50$\pm$0.5    &  51.70$\pm$1.2  &  [44] \\
  $^{40}$Ca+$^{48}$Ca     &   1.2   & 400 &   10.08    &  57.14    &   -3.96    &   53.18   &   7.80$\pm$0.3    &  51.30$\pm$1.0  &  [44] \\
  $^{48}$Ca+$^{48}$Ca     &   1.4   & 400 &   10.38    &  55.50    &   -3.75    &   51.75   &   10.38           &  51.70          &  [45] \\
  \\
  $^{40}$Ca+$^{46}$Ti     &   1.048   & 440 &   9.91     &  63.94    &   -4.57    &   59.37   &   9.92$\pm$0.08    &  58.03$\pm$0.73  &  [47] \\
  $^{40}$Ca+$^{48}$Ti     &   1.095   & 440 &   10.00    &  63.36    &   -4.44    &   58.92   &   9.97$\pm$0.07    &  58.17$\pm$0.62  &  [47] \\
  $^{40}$Ca+$^{50}$Ti     &   1.143   & 440 &   10.07    &  62.92    &   -4.43    &   58.49   &   10.05$\pm$0.07   &  58.71$\pm$0.61  &  [47] \\
  \\
  $^{32}$S+$^{58}$Ni     &   1.045   & 448 &   9.95      &  64.84    &   -4.61    &   60.23   &   8.60$\pm$0.9    &  59.80$\pm$1.4  &  [43] \\
                         &           &     &             &           &            &           &   8.50$\pm$0.3    &  59.50          &  [48] \\
  $^{32}$S+$^{64}$Ni     &   1.182   & 448 &   10.16     &  63.50    &   -4.38    &   59.12   &   8.80$\pm$0.5    &  58.10$\pm$0.7  &  [43] \\
  $^{34}$S+$^{58}$Ni     &   1.090   & 448 &   10.06     &  64.13    &   -4.48    &   59.65   &   7.50$\pm$0.9    &  58.40$\pm$1.4  &  [43] \\
  $^{34}$S+$^{64}$Ni     &   1.227   & 448 &   10.25     &  62.95    &   -4.34    &   58.61   &   8.90$\pm$0.6    &  57.20$\pm$0.6  &  [43] \\
  $^{36}$S+$^{58}$Ni     &   1.136   & 448 &   10.16     &  63.50    &   -4.39    &   59.11   &   7.50$\pm$0.6    &  58.00$\pm$1.1  &  [43] \\
  $^{36}$S+$^{64}$Ni     &   1.273   & 448 &   10.34     &  62.39    &   -4.27    &   58.12   &   8.80$\pm$0.6    &  56.70$\pm$1.0  &  [43] \\
  \\
  $^{40}$Ar+$^{58}$Ni     &   1.130   & 504 &   10.24    &  70.88    &   -4.96    &   65.92   &    &  65.30$\pm$0.5  &  [49] \\
  $^{40}$Ar+$^{60}$Ni     &   1.174   & 504 &   10.31    &  70.40    &   -4.86    &   65.54   &    &  65.50$\pm$0.6  &  [49] \\
  $^{40}$Ar+$^{62}$Ni     &   1.217   & 504 &   10.37    &  70.00    &   -4.82    &   65.18   &    &  65.10$\pm$0.6  &  [49] \\
  $^{40}$Ar+$^{64}$Ni     &   1.260   & 504 &   10.43    &  69.59    &   -4.77    &   64.82   &    &  63.90$\pm$0.5  &  [49] \\
  \\
  $^{40}$Ca+$^{58}$Ni     &   1.042   & 560 &   10.15    &  79.45    &   -5.65    &   73.80   &   10.20   &  73.36  &  [36] \\
  $^{40}$Ca+$^{62}$Ni     &   1.125   & 560 &   10.29    &  78.38    &   -5.48    &   72.90   &   10.35   &  72.30  &  [36] \\
  \\
  $^{48}$Ti+$^{58}$Ni     &   1.12   & 616 &   10.4     &  85.30    &   -6.00    &   79.30   &   9.8$\pm$0.3    &  78.8$\pm$0.3  &  [50] \\
  $^{48}$Ti+$^{60}$Ni     &   1.16   & 616 &   10.47    &  84.73    &   -5.89    &   78.84   &   10.0$\pm$0.3   &  77.3$\pm$0.3  &  [50] \\
  $^{48}$Ti+$^{64}$Ni     &   1.24   & 616 &   10.60    &  83.69    &   -5.72    &   77.97   &   10.2$\pm$0.3   &  76.7$\pm$0.3  &  [50] \\
  $^{46}$Ti+$^{64}$Ni     &   1.2    & 616 &   10.52    &  84.33    &   -5.84    &   78.49   &   9.7$\pm$0.2    &  76.9$\pm$0.1  &  [51] \\
  $^{50}$Ti+$^{60}$Ni     &   1.2    & 616 &   10.55    &  84.09    &   -5.77    &   78.32   &   9.8$\pm$0.2    &  77.1$\pm$0.1  &  [51] \\
  \\
  $^{58}$Ni+$^{58}$Ni     &   1.071   & 784 &   10.55    &  107.03    &   -7.62    &   99.41   &   8.30   &  97.90  &  [52] \\
  $^{58}$Ni+$^{64}$Ni     &   1.178   & 784 &   10.75    &  105.04    &   -7.33    &   97.71   &   8.20   &  96.00  &  [52] \\
  $^{64}$Ni+$^{64}$Ni     &   1.286   & 784 &   10.94    &  103.21    &   -7.04    &   96.17   &   8.60   &  93.50  &  [52] \\
  \hline
\end{tabular}
\end{center}
\newpage
Table 2. The calculated barrier positions $R_B$ and heights $V_B$
for different fusion reactions using Bass 80 potential. The nuclear
and Coulomb potentials in $r=R_B$ have also been reported. The
systems are listed with respect to their increasing $Z_{1}Z_{2}$
values.

\begin{center}
\begin{tabular}{c c c c c c c}
  \hline
  \hline
  Reaction & N/Z & $Z_{1}Z_{2}$ &  $R_B$   &  $V_C$   &  $V_N$   &  $V_B$ \\
  \hline
  %\cline{4-10}
  $^{12}$C+$^{28}$Si     &   1      & 84 &   8.45    &  14.31    &   -1.17    &   13.14   \\
  $^{12}$C+$^{29}$Si     &   1.05   & 84 &   8.51    &  14.21    &   -1.16    &   13.05   \\
  $^{12}$C+$^{30}$Si     &   1.1    & 84 &   8.57    &  14.11    &   -1.14    &   12.97   \\
  \\
  $^{16}$O+$^{24}$Mg     &   1      & 96 &   8.50    &  16.26    &   -1.33    &   14.93   \\
                         &          &    &           &           &            &           \\
  $^{16}$O+$^{26}$Mg     &   1.1    & 96 &   8.65    &  15.98    &   -1.26    &   14.72   \\
  $^{18}$O+$^{24}$Mg     &   1.1    & 96 &   8.68    &  15.93    &   -1.27    &   14.65   \\
  \\
  $^{16}$O+$^{28}$Si     &   1      & 112 &  8.63    &  18.69    &   -1.51    &   17.18   \\
  $^{16}$O+$^{29}$Si     &   1.045  & 112 &  8.69    &  18.56    &   -1.49    &   17.07   \\
  $^{16}$O+$^{30}$Si     &   1.090  & 112 &  8.75    &  18.42    &   -1.46    &   16.96   \\
  $^{18}$O+$^{28}$Si     &   1.090  & 112 &  8.81    &  18.31    &   -1.45    &   16.86   \\
  \\
  $^{24}$Mg+$^{32}$S     &   1        & 192 &   9.03    &  30.61    &   -2.43    &   28.18 \\
  $^{24}$Mg+$^{34}$S     &   1.071    & 192 &   9.14    &  30.25    &   -2.37    &   27.88 \\
  $^{26}$Mg+$^{32}$S     &   1.071    & 192 &   9.17    &  30.14    &   -2.34    &   27.80 \\
  $^{26}$Mg+$^{34}$S     &   1.143    & 192 &   9.28    &  29.79    &   -2.29    &   27.50 \\
  \\
  $^{28}$Si+$^{28}$Si     &   1        & 196 &   9.04    &  31.22    &   -2.48    &   28.74 \\
  $^{28}$Si+$^{30}$Si     &   1.071    & 196 &   9.16    &  30.81    &   -2.41    &   28.40 \\
  $^{30}$Si+$^{30}$Si     &   1.143    & 196 &   9.29    &  30.38    &   -2.32    &   28.06 \\
  \\
  $^{28}$Si+$^{58}$Ni     &   1.048   & 392 &   9.79     &  57.66    &   -4.44    &   53.22  \\
  $^{28}$Si+$^{62}$Ni     &   1.143   & 392 &   9.94     &  56.79    &   -4.27    &   52.52  \\
  $^{28}$Si+$^{64}$Ni     &   1.190   & 392 &   10.01     &  56.40    &   -4.20    &  52.19  \\
  $^{30}$Si+$^{58}$Ni     &   1.095   & 392 &   9.92     &  56.91    &   -4.29    &   52.62  \\
  $^{30}$Si+$^{62}$Ni     &   1.190   & 392 &   10.07     &  56.06    &   -4.12    &   51.94 \\
  $^{30}$Si+$^{64}$Ni     &   1.238   & 392 &   10.14     &  56.68    &   -4.06    &   51.61 \\
  \hline
\end{tabular}
\end{center}

\newpage
Table 2. (Continued)

\begin{center}
\begin{tabular}{c c c c c c c }
  \hline
  \hline
  Reaction & N/Z & $Z_{1}Z_{2}$& $R_B$   &  $V_C$   &  $V_N$   &  $V_B$  \\
  \hline
  %\cline{4-10}
  $^{40}$Ca+$^{40}$Ca     &   1     & 400 &   9.68     &  59.51    &   -4.63    &  54.88   \\
  $^{40}$Ca+$^{44}$Ca     &   1.1   & 400 &   9.87     &  58.36    &   -4.43    &   53.93  \\
  $^{40}$Ca+$^{48}$Ca     &   1.2   & 400 &   10.05     &  57.32    &   -4.24    &   53.08 \\
  $^{48}$Ca+$^{48}$Ca     &   1.4   & 400 &   10.42     &  55.28    &   -3.88    &   51.40 \\
  \\
  $^{40}$Ca+$^{46}$Ti     &   1.048   & 440 &   9.86     &  64.26    &   -4.96    &   59.30 \\
  $^{40}$Ca+$^{48}$Ti     &   1.095   & 440 &   9.95     &  63.69    &   -4.85    &   58.84 \\
  $^{40}$Ca+$^{50}$Ti     &   1.143   & 440 &   10.04    &  63.12    &   -4.72    &   58.40 \\
  \\
  $^{32}$S+$^{58}$Ni     &   1.045   & 448 &   9.90      &  65.17    &   -5.02    &   60.15 \\
  $^{32}$S+$^{64}$Ni     &   1.182   & 448 &   10.12     &  63.76    &   -4.76    &   59.00 \\
  $^{34}$S+$^{58}$Ni     &   1.090   & 448 &   10.02     &  64.39    &   -4.86    &   59.53 \\
  $^{34}$S+$^{64}$Ni     &   1.227   & 448 &   10.24     &  63.01    &   -4.60    &   58.41 \\
  $^{36}$S+$^{58}$Ni     &   1.136   & 448 &   10.13     &  63.69    &   -4.73    &   58.96 \\
  $^{36}$S+$^{64}$Ni     &   1.273   & 448 &   10.35     &  62.33    &   -4.48    &   57.85 \\
  \\
  $^{40}$Ar+$^{58}$Ni     &   1.130   & 504 &   10.21    &  71.09    &   -5.33    &   65.76 \\
  $^{40}$Ar+$^{60}$Ni     &   1.174   & 504 &   10.29    &  70.54    &   -5.20    &   65.34 \\
  $^{40}$Ar+$^{62}$Ni     &   1.217   & 504 &   10.36    &  70.05    &   -5.13    &   64.93 \\
  $^{40}$Ar+$^{64}$Ni     &   1.260   & 504 &   10.43    &  69.60    &   -5.06    &   64.54 \\
  \\
  $^{40}$Ca+$^{58}$Ni     &   1.042   & 560 &   10.10    &  79.85    &   -6.15    &   73.70 \\
  $^{40}$Ca+$^{62}$Ni     &   1.125   & 560 &   10.25    &  78.69    &   -5.93    &   72.76 \\
  \\
  $^{48}$Ti+$^{58}$Ni     &   1.12   & 616 &   10.37    &  85.55    &   -6.45    &   79.30  \\
  $^{48}$Ti+$^{60}$Ni     &   1.16   & 616 &   10.45    &  84.90    &   -6.30    &   78.84  \\
  $^{48}$Ti+$^{64}$Ni     &   1.24   & 616 &   10.59    &  83.77    &   -6.13    &   77.97  \\
  $^{46}$Ti+$^{64}$Ni     &   1.2    & 616 &   10.50    &  84.49    &   -6.27    &   78.49  \\
  $^{50}$Ti+$^{60}$Ni     &   1.2    & 616 &   10.54    &  84.16    &   -6.14    &   78.32  \\
  \\
  $^{58}$Ni+$^{58}$Ni     &   1.071   & 784 &   10.51    &  107.43    &   -8.25    &   99.18\\
  $^{58}$Ni+$^{64}$Ni     &   1.178   & 784 &   10.74    &  105.14    &   -7.77    &   97.36\\
  $^{64}$Ni+$^{64}$Ni     &   1.286   & 784 &   10.96    &  103.02    &   -7.41    &   95.61\\
  \hline
\end{tabular}
\end{center}
\newpage
Table 3. The calculated barrier positions $R_B$ and heights $V_B$
for different fusion reactions using Prox. DP potential. The nuclear
and Coulomb potentials in $r=R_B$ have also been reported. The
systems are listed with respect to their increasing $Z_{1}Z_{2}$
values.

\begin{center}
\begin{tabular}{c c c c c c c}
  \hline
  \hline
  Reaction & N/Z & $Z_{1}Z_{2}$& $R_B$   &  $V_C$   &  $V_N$   &  $V_B$ \\
  \hline
  %\cline{4-10}
  $^{12}$C+$^{28}$Si     &   1      & 84 &   8.36    &  14.46    &   -1.31    &   13.15   \\
  $^{12}$C+$^{29}$Si     &   1.05   & 84 &   8.43    &  14.34    &   -1.30    &   13.04   \\
  $^{12}$C+$^{30}$Si     &   1.1    & 84 &   8.50    &  14.22    &   -1.30    &   12.92   \\
  \\
  $^{16}$O+$^{24}$Mg     &   1      & 96 &   8.37    &  16.51    &   -1.50    &   15.01   \\
                         &          &    &           &           &            &           \\
  $^{16}$O+$^{26}$Mg     &   1.1    & 96 &   8.53    &  16.20    &   -1.48    &   14.72   \\
  $^{18}$O+$^{24}$Mg     &   1.1    & 96 &   8.60    &  16.07    &   -1.47    &   14.60   \\
  \\
  $^{16}$O+$^{28}$Si     &   1      & 112 &  8.48    &  19.01    &   -1.72    &   17.29   \\
  $^{16}$O+$^{29}$Si     &   1.045  & 112 &  8.56    &  18.83    &   -1.69    &   17.14   \\
  $^{16}$O+$^{30}$Si     &   1.090  & 112 &  8.63    &  18.68    &   -1.68    &   17.00   \\
  $^{18}$O+$^{28}$Si     &   1.090  & 112 &  8.72    &  18.48    &   -1.66    &   16.82   \\
  \\
  $^{24}$Mg+$^{32}$S     &   1        & 192 &   8.84    &  31.27    &   -2.76    &   28.51 \\
  $^{24}$Mg+$^{34}$S     &   1.071    & 192 &   8.98    &  30.78    &   -2.68    &   28.10 \\
  $^{26}$Mg+$^{32}$S     &   1.071    & 192 &   9.02    &  30.65    &   -2.66    &   27.99 \\
  $^{26}$Mg+$^{34}$S     &   1.143    & 192 &   9.16    &  30.18    &   -2.58    &   27.60 \\
  \\
  $^{28}$Si+$^{28}$Si     &   1       & 196 &    8.85    &  31.89    &   -2.80    &   29.08 \\
  $^{28}$Si+$^{30}$Si     &   1.071    & 196 &   9.00    &  31.35    &   -2.74    &   28.61 \\
  $^{30}$Si+$^{30}$Si     &   1.143    & 196 &   9.16    &  30.81    &   -2.64    &   28.17 \\
  \\
  $^{28}$Si+$^{58}$Ni     &   1.048   & 392 &   9.62     &  58.67    &   -4.73    &   53.94  \\
  $^{28}$Si+$^{62}$Ni     &   1.143   & 392 &   9.78     &  57.71    &   -4.57    &   53.14  \\
  $^{28}$Si+$^{64}$Ni     &   1.190   & 392 &   9.85     &  57.30    &   -4.52    &   52.78  \\
  $^{30}$Si+$^{58}$Ni     &   1.095   & 392 &   9.78     &  57.71    &   -4.57    &   53.14  \\
  $^{30}$Si+$^{62}$Ni     &   1.190   & 392 &   9.94     &  56.78    &   -4.41    &   52.37  \\
  $^{30}$Si+$^{64}$Ni     &   1.238   & 392 &   10.01     &  56.39    &   -4.37    &   52.01 \\
  \hline
\end{tabular}
\end{center}

\newpage
Table 3. (Continued)

\begin{center}
\begin{tabular}{c c c c c c c}
  \hline
  \hline
  Reaction & N/Z & $Z_{1}Z_{2}$& $R_B$   &  $V_C$   &  $V_N$   &  $V_B$ \\
  \hline
  %\cline{4-10}
  $^{40}$Ca+$^{40}$Ca     &   1     & 400 &   9.50     &  60.62    &   -4.95    &   55.67   \\
  $^{40}$Ca+$^{44}$Ca     &   1.1   & 400 &   9.72     &  59.25    &   -4.74    &   54.51   \\
  $^{40}$Ca+$^{48}$Ca     &   1.2   & 400 &   9.92     &  58.05    &   -4.54    &   53.51   \\
  $^{48}$Ca+$^{48}$Ca     &   1.4   & 400 &   10.33     & 55.75    &   -4.23    &   51.52   \\
  \\
  $^{40}$Ca+$^{46}$Ti     &   1.048   & 440 &   9.70     &  65.31    &   -5.24    &   60.07 \\
  $^{40}$Ca+$^{48}$Ti     &   1.095   & 440 &   9.80    &  64.64    &   -5.13    &   59.51  \\
  $^{40}$Ca+$^{50}$Ti     &   1.143   & 440 &   9.90    &  64.00    &   -5.00    &   59.00  \\
  \\
  $^{32}$S+$^{58}$Ni     &   1.045   & 448 &   9.74     &  66.23    &   -5.25    &   60.97  \\
  $^{32}$S+$^{64}$Ni     &   1.182   & 448 &   9.97     &  64.69    &   -5.02    &   59.67  \\
  $^{34}$S+$^{58}$Ni     &   1.090   & 448 &   9.88     &  65.29    &   -5.10    &   60.18  \\
  $^{34}$S+$^{64}$Ni     &   1.227   & 448 &   10.11     &  63.80    &   -4.88    &   58.92 \\
  $^{36}$S+$^{58}$Ni     &   1.136   & 448 &   10.01     &  64.44    &   -4.97    &   59.47 \\
  $^{36}$S+$^{64}$Ni     &   1.273   & 448 &   10.24     &  62.99    &   -4.76    &   58.23 \\
  \\
  $^{40}$Ar+$^{58}$Ni     &   1.130   & 504 &   10.08    &  71.99    &   -5.59    &   66.40 \\
  $^{40}$Ar+$^{60}$Ni     &   1.174   & 504 &   10.17    &  71.35    &   -5.43    &   65.92 \\
  $^{40}$Ar+$^{62}$Ni     &   1.217   & 504 &   10.24    &  70.87    &   -5.40    &   65.47 \\
  $^{40}$Ar+$^{64}$Ni     &   1.260   & 504 &   10.32    &  70.31    &   -5.28    &   65.03 \\
  \\
  $^{40}$Ca+$^{58}$Ni     &   1.042   & 560 &   9.93    &   81.20    &   -6.46    &   74.74 \\
  $^{40}$Ca+$^{62}$Ni     &   1.125   & 560 &   10.09    &  79.91    &   -6.24    &   73.67 \\
  \\
  $^{48}$Ti+$^{58}$Ni     &   1.12   & 616 &   10.23     &  86.70    &   -6.70    &   80.000\\
  $^{48}$Ti+$^{60}$Ni     &   1.16   & 616 &   10.31    &  86.03    &   -6.60    &   79.43  \\
  $^{48}$Ti+$^{64}$Ni     &   1.24   & 616 &   10.47    &  84.71    &   -6.34    &   78.37  \\
  $^{46}$Ti+$^{64}$Ni     &   1.2    & 616 &   10.37    &  85.53    &   -6.47    &   79.06  \\
  $^{50}$Ti+$^{60}$Ni     &   1.2    & 616 &   10.41    &  85.20    &   -6.43    &   78.77  \\
  \\
  $^{58}$Ni+$^{58}$Ni     &   1.071   & 784 &   10.34    &  109.17    &   -8.62    &   100.55 \\
  $^{58}$Ni+$^{64}$Ni     &   1.178   & 784 &   10.59    &  106.60    &   -8.08    &   98.52  \\
  $^{64}$Ni+$^{64}$Ni     &   1.286   & 784 &   10.83    &  104.24    &   -7.67    &   96.57  \\
  \hline
\end{tabular}
\end{center}
\newpage
Table 4. The calculated barrier positions $R_B$ and heights $V_B$
for different fusion reactions using Prox. 2010 potential. The
nuclear and Coulomb potentials in $r=R_B$ have also been reported.
The systems are listed with respect to their increasing $Z_{1}Z_{2}$
values.

\begin{center}
\begin{tabular}{c c c c c c c}
  \hline
  \hline
  Reaction & N/Z & $Z_{1}Z_{2}$& $R_B$   &  $V_C$   &  $V_N$   &  $V_B$  \\
  \hline
  %\cline{4-10}
  $^{12}$C+$^{28}$Si     &   1      & 84 &   8.29    &  14.58    &   -1.32    &   13.26   \\
  $^{12}$C+$^{29}$Si     &   1.05   & 84 &   8.35    &  14.48    &   -1.30    &   13.18   \\
  $^{12}$C+$^{30}$Si     &   1.1    & 84 &   8.41    &  14.37    &   -1.27    &   13.10   \\
  \\
  $^{16}$O+$^{24}$Mg     &   1      & 96 &   8.34    &  16.52    &   -1.48    &   15.08  \\
  $^{16}$O+$^{26}$Mg     &   1.1    & 96 &   8.47    &  16.32    &   -1.44    &   14.88  \\
  $^{18}$O+$^{24}$Mg     &   1.1    & 96 &   8.52    &  16.22    &   -1.42    &   14.80  \\
  \\
  $^{16}$O+$^{28}$Si     &   1      & 112 &  8.46    &  19.05    &   -1.69    &   17.36  \\
  $^{16}$O+$^{29}$Si     &   1.045  & 112 &  8.52    &  18.92    &   -1.67    &   17.25  \\
  $^{16}$O+$^{30}$Si     &   1.090  & 112 &  8.58    &  18.79    &   -1.64    &   17.15  \\
  $^{18}$O+$^{28}$Si     &   1.090  & 112 &  8.64    &  18.65    &   -1.62    &   17.03  \\
  \\
  $^{24}$Mg+$^{32}$S     &   1        & 192 &   8.88    &  31.13    &   -2.62    &   28.5   \\
  $^{24}$Mg+$^{34}$S     &   1.071    & 192 &   8.99    &  30.75    &   -2.55    &   28.20  \\
  $^{26}$Mg+$^{32}$S     &   1.071    & 192 &   9.01    &  30.68    &   -2.56    &   28.12  \\
  $^{26}$Mg+$^{34}$S     &   1.143    & 192 &   9.11    &  30.34    &   -2.50    &   27.84  \\
  \\
  $^{28}$Si+$^{28}$Si     &   1        & 196 &   8.89    &  31.74    &   -2.67    &   29.07 \\
  $^{28}$Si+$^{30}$Si     &   1.071    & 196 &   9.01    &  31.31    &   -2.59    &   28.72 \\
  $^{30}$Si+$^{30}$Si     &   1.143    & 196 &   9.12    &  30.44    &   -2.53    &   28.41 \\
  \\
  $^{28}$Si+$^{58}$Ni     &   1.048   & 392 &   9.68     &  58.30    &   -4.49    &   53.81 \\
  $^{28}$Si+$^{62}$Ni     &   1.143   & 392 &   9.80     &  57.59    &   -4.43    &   53.16 \\
  $^{28}$Si+$^{64}$Ni     &   1.190   & 392 &   9.86     &  57.24    &   -4.36    &   52.88 \\
  $^{30}$Si+$^{58}$Ni     &   1.095   & 392 &   9.79     &  57.65    &   -4.43    &   53.22 \\
  $^{30}$Si+$^{62}$Ni     &   1.190   & 392 &   9.92     &  56.89    &   -4.28    &   52.61 \\
  $^{30}$Si+$^{64}$Ni     &   1.238   & 392 &   9.97     &  56.61    &   -4.26    &   52.35 \\
  \hline
\end{tabular}
\end{center}

\newpage
Table 4. (Continued)

\begin{center}
\begin{tabular}{c c c c c c c}
  \hline
  \hline
  Reaction & N/Z & $Z_{1}Z_{2}$& $R_B$   &  $V_C$   &  $V_N$   &  $V_B$  \\
  \hline
  %\cline{4-10}
  $^{40}$Ca+$^{40}$Ca     &   1     & 400 &   9.57     &  60.18    &   -4.70    &   55.48   \\
  $^{40}$Ca+$^{44}$Ca     &   1.1   & 400 &   9.75     &  59.07    &   -4.52    &   54.55   \\
  $^{40}$Ca+$^{48}$Ca     &   1.2   & 400 &   9.90     &  58.17    &   -4.39    &   53.78   \\
  $^{48}$Ca+$^{48}$Ca     &   1.4   & 400 &   10.18    &  56.57    &   -4.17    &   52.40   \\
  \\
  $^{40}$Ca+$^{46}$Ti     &   1.048   & 440 &   9.76    &  64.91    &   -4.96    &   59.95  \\
  $^{40}$Ca+$^{48}$Ti     &   1.095   & 440 &   9.84    &  64.38    &   -4.88    &   59.50  \\
  $^{40}$Ca+$^{50}$Ti     &   1.143   & 440 &   9.91    &  63.92    &   -4.83    &   59.09  \\
  \\
  $^{32}$S+$^{58}$Ni     &   1.045   & 448 &   9.80     &  65.82    &   -5.02    &   60.08  \\
  $^{32}$S+$^{64}$Ni     &   1.182   & 448 &   9.99     &  64.57    &   -4.82    &   59.75  \\
  $^{34}$S+$^{58}$Ni     &   1.090   & 448 &   9.90     &  65.16    &   -4.96    &   60.20  \\
  $^{34}$S+$^{64}$Ni     &   1.227   & 448 &   10.08     &  63.99    &   -4.78    &   59.21 \\
  $^{36}$S+$^{58}$Ni     &   1.136   & 448 &   10.00     &  64.50    &   -4.85   &   59.65  \\
  $^{36}$S+$^{64}$Ni     &   1.273   & 448 &   10.18     &  63.36    &   -4.64    &   58.72 \\
  \\
  $^{40}$Ar+$^{58}$Ni     &   1.130   & 504 &   10.10    &  71.85    &   -5.34    &   66.51 \\
  $^{40}$Ar+$^{60}$Ni     &   1.174   & 504 &   10.16    &  71.43    &   -5.29    &   66.14 \\
  $^{40}$Ar+$^{62}$Ni     &   1.217   & 504 &   10.22    &  71.01    &   -5.22    &   65.79 \\
  $^{40}$Ar+$^{64}$Ni     &   1.260   & 504 &   10.28    &  70.59    &   -5.13    &   65.46 \\
  \\
  $^{40}$Ca+$^{58}$Ni     &   1.042   & 560 &   10.02    &  80.47    &   -6.02    &   74.45 \\
  $^{40}$Ca+$^{62}$Ni     &   1.125   & 560 &   10.15    &  79.44    &   -5.88    &   73.56 \\
  \\
  $^{48}$Ti+$^{58}$Ni     &   1.12   & 616 &   10.29    &  86.20    &   -6.25    &   79.95   \\
  $^{48}$Ti+$^{60}$Ni     &   1.16   & 616 &   10.35    &  85.69    &   -6.20    &   79.49   \\
  $^{48}$Ti+$^{64}$Ni     &   1.24   & 616 &   10.47    &  84.71    &   -6.04    &   78.67   \\
  $^{46}$Ti+$^{64}$Ni     &   1.2    & 616 &   10.39    &  85.37    &   -6.18    &   79.19   \\
  $^{50}$Ti+$^{60}$Ni     &   1.2    & 616 &   10.42    &  85.12    &   -6.14    &   78.98   \\
  \\
  $^{58}$Ni+$^{58}$Ni     &   1.071   & 784 &   10.47    &  107.82    &   -7.74    &   100.08 \\
  $^{58}$Ni+$^{64}$Ni     &   1.178   & 784 &   10.66    &  105.90    &   -7.48    &   98.42  \\
  $^{64}$Ni+$^{64}$Ni     &   1.286   & 784 &   10.84    &  104.13    &   -7.21    &   96.92  \\
  \hline
\end{tabular}
\end{center}
\newpage
Table 5. The calculated values for fitting parameters $\alpha$,
$\beta$, $\gamma$ and $c$, which are based on the Eqs. (14), (15),
(16) and (20), respectively, for AW 95, Bass 80, Prox. DP and Prox.
2010 potentials.

\begin{center}
\begin{tabular}{c c c c c c}
  \hline
  \hline
  Proximity-type potential   &    $\alpha$    &  $\beta$   &  $\gamma$ &    c($E^0_{c.m.}$=1.125)   &    c($E^0_{c.m.}$=1.375)  \\
  \hline
  AW 95      &   16.18   &   -14.31     &   31.45  &   155.83     &  73.57    \\
  Bass 80    &   18.26   &   -15.33     &   41.82  &   170.21     &  80.02    \\
  Prox. DP   &   20.78   &   -18.11     &   38.56  &   192.58     &  94.06    \\
  Prox. 2010 &   16.57   &   -13.63     &   28.80  &   146.33     &  69.78    \\
  \hline
\end{tabular}
\end{center}

\newpage
\begin{figure}
\begin{center}
\includegraphics{fig.}
\end{center}
\vspace{15cm} \caption{}
\end{figure}

\newpage
\begin{figure}
\begin{center}
\includegraphics{fig.}
\end{center}
\vspace{16cm} \caption{}
\end{figure}

\newpage
\begin{figure}
\begin{center}
\includegraphics{fig.}
\end{center}
\vspace{14cm} \caption{}
\end{figure}

\newpage
\begin{figure}
\begin{center}
\includegraphics{fig.}
\end{center}
\vspace{14cm} \caption{}
\end{figure}

\newpage
\begin{figure}
\begin{center}
\includegraphics{fig.}
\end{center}
\vspace{14cm} \caption{}
\end{figure}

\newpage
\begin{figure}
\begin{center}
\includegraphics{fig.}
\end{center}
\vspace{16cm} \caption{}
\end{figure}

\newpage
\begin{figure}
\begin{center}
\includegraphics{fig.}
\end{center}
\vspace{16cm} \caption{}
\end{figure}

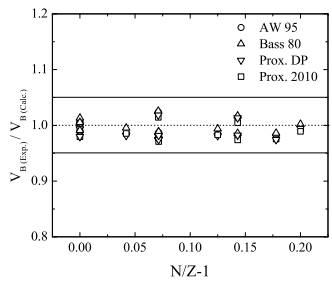
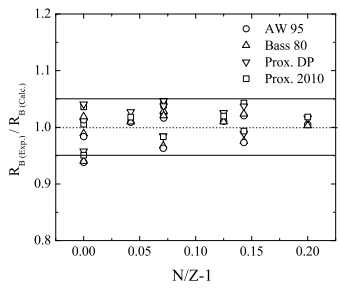
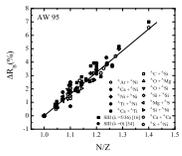
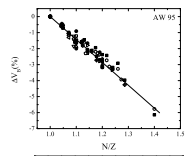
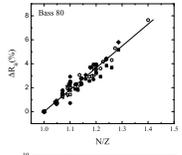
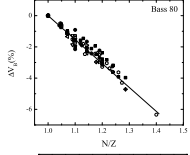
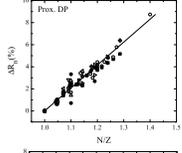
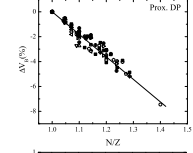
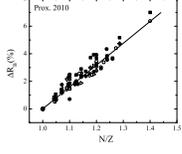
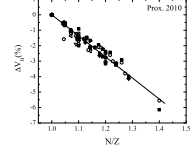
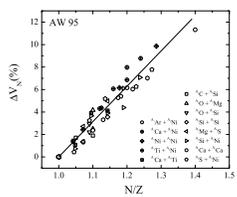
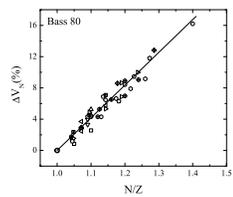
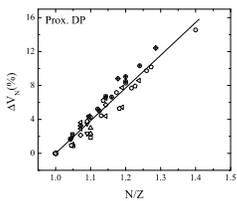
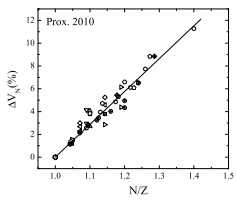
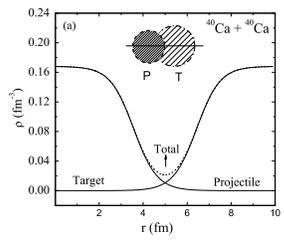
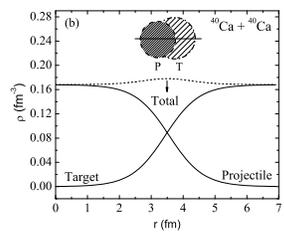
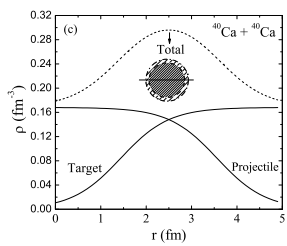
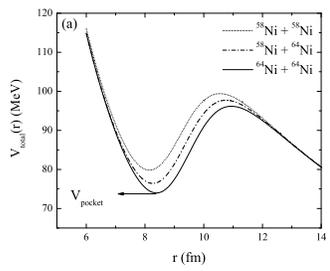
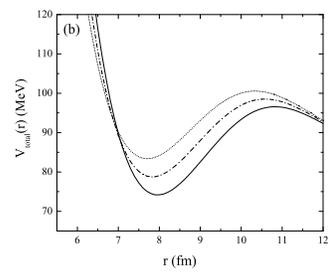
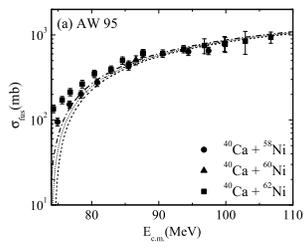
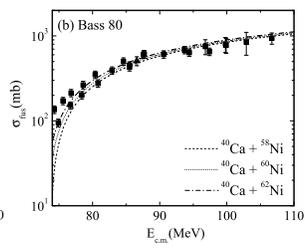
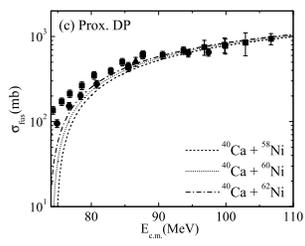
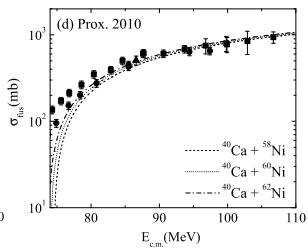
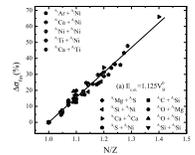
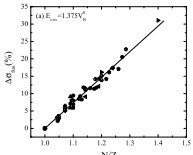
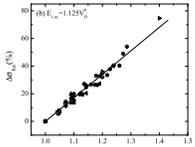
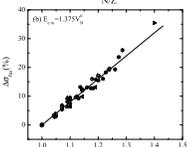
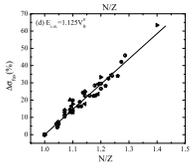
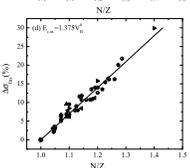
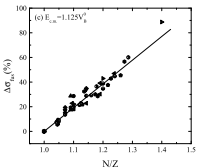
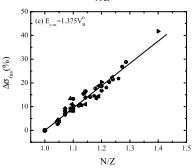
\end{document}